\documentclass[         %
aps,                    %  American Physical Society
prd,                    %  Physical Review D
showpacs,               %  Displays PACS after abstract
nofootinbib,            %  Does not treat footnotes as references
preprintnumbers,        %
floatfix]               %  Fixes float errors
{revtex4}               %  REVTEX 4 Package used
\usepackage{graphicx}
\usepackage[dvips]{epsfig}
\usepackage{dcolumn}%Align table columns on decimal point
\usepackage{bm}% bold math
\usepackage{fancybox}
\begin{document}

\title{Unparticle effects on top quark rare decays}

\author{T. M. Aliev\footnote{Permanent address:
Institute of Physics, Baku, Azerbaijan}}
\email{taliev@metu.edu.tr}
\affiliation{Department of Physics, Faculty of Arts and Sciences, Middle East
Technical University, Ankara, Turkey}
\author{O. Cakir}
\email{ocakir@science.ankara.edu.tr}
\affiliation{Physics Department CERN, 1211 Geneva 23, Switzerland}
\affiliation{Department of Physics, Faculty of Sciences, 
Ankara University, 06100 Tandogan, Ankara, Turkey}
\author{K. O. Ozansoy}
\email{oozansoy@physics.wisc.edu}
\affiliation{Department of Physics, Faculty of Sciences, Ankara University, 06100 Tandogan, Ankara, Turkey}

\date{\today}

\begin{abstract}
In this work we study the flavor changing neutral current(FCNC) 
decays of the top quark, $t\to c\gamma$ and $t\to c g$,
in the framework of the unparticle physics. The Standard 
Model predictions for the branching ratios
of these decays are about $\sim 5\times 10^{-14}$,
and $\sim 1\times 10^{-12}$, respectively. The parameter space  
of $\lambda$, $\Lambda$, and $d$  is obtained by taking
into account the SM predictions and the results of
the simulation performed by the ATLAS  Collaboration 
for the branching ratios of $t\to c\gamma$ and $t\to c g$ decays.
\end{abstract}

\pacs{12.60.-i, 14.80.-j}
%\keywords{unparticle sector, top quark rare decays}

\maketitle

\section{Introduction}

After the Large Hadron Collider(LHC) has been launched very recently,
next decade will be a stage for a better understanding of the nature of the 
properties of, and the interactions among the elementary particles at
TeV scale. On the one hand, LHC is expected to give a perfect understanding
of the electroweak symmetry breaking of the Standard Model(SM) which is 
expressed through the Higgs mechanism. On the other hand, diversity of the 
new physics scenarios will be sought at the LHC. Having a mass about the 
electroweak scale and being the heaviest particle in the SM the top quark 
is one of the beacons of the LHC to shed light on the riddles of the
electroweak symmetry breaking, and to explore the new physics effects at TeV 
scale. Importance of the top quark searches at the LHC has been 
concisely reviewed by Ref.s~ \cite{Beneke:2000hk}, \cite{Han:2008xb}, and
\cite{Bernreuther:2008ju}. 

Since there will be huge amount production ($\sim 80$ million pair, and $\sim 30$
million single, Ref.~\cite{Han:2008xb}) of the top quark at the LHC, 
one can predict 
that the interest of the top quark studies will be grown. There are 
two possibilities to seek for the new physics effects through 
the top quark decays, one is the decays via charged currents 
and the other is the decays via  neutral currents.

In the SM, the top quark mainly decays to a $W^+$ boson and a 
quark $q$, ($q=d,s,b$ ), Ref.~\cite{Yao:2006px}. 
As a very important remark to explore the several new
physics predictions, besides those charge current decays of the top quark
there is no tree level decay of the top quark through neutral currents
in the SM. New physics searches via the top quark decays have been extensively 
analysed in the literature (see Refs~\cite{Chakraborty:2003iw, Eilam:1990zc, 
Han:2008xb, AguilarSaavedra:2004wm, Frank:2005vd}, 
and references there in). 

The flavor changing neutral current(FCNC) decays of the top quark 
are highly suppresed in the SM (namely, the branching ratios 
for $t\to q Z,  q \gamma, q g$ are predicted about from 
$\sim 10^{-15}$ to $\sim 10^{-11}$) due to the
Glashow-Iliopoulos-Maiani(GIM) mechanism, 
eg. Ref.s~\cite{Eilam:1990zc, Mele:1998ag,AguilarSaavedra:2004wm}.
Note that recently, 
simulation performed by
the ATLAS Colloboration gives  
upper bound $\sim 10^{-5}$ on $t\to q\gamma(g)$ decay at $95\%$ C.L.,
\cite{Carvalho:2007yi, Han:1995pk}. There are many beyond 
the SM proposals to predict FCNC decays of the top
quark ( a good literature is given by 
Ref.s~\cite{Frank:2005vd, Bernreuther:2008ju}).
For example in the (minimal, or left-right)supersymmetric standard 
model scenario, Ref.s~\cite{Li:1993mg},~\cite{deDivitiis:1997sh}, 
\cite{Lopez:1997xv}, or in the littlest Higgs
model scenerio, Ref.~\cite{HongSheng:2007ve}, or
in the left-right supersymmetric model, Ref.~\cite{Frank:2005vd} 
those ratios are found about $10^{-3}-10^{-6}$.

One of the most interesting and mind-bending recent new physics 
scenarios is the unparticle physics which is proposed
by Georgi, Ref.~\cite{Georgi:2007ek, Georgi:2007si}. 
According to unparticle physics proposal given by Georgi, 
if there is a conformal symmetry in nature it must be broken 
at a very high energy scale which is above the current energy scale of 
the colliders. Considering the idea of 
Ref.~\cite{Banks:1981nn}, in Ref.~\cite{Georgi:2007ek}, 
the scale invariant sector is presented by a set of the 
Banks-Zaks operators ${\cal O}_{BZ}$, and defined at 
the very high energy scale. Interactions of BZ 
operators ${\cal O}_{BZ}$ with the SM operators 
${\cal O}_{SM}$ are expressed by the exchange of
particles with a very high energy mass scale
${\cal M}_{\cal U}^k$ in the following form

\begin{eqnarray}
\label{1}
 \frac{1}{{\cal M}_{\cal U}^k}{O}_{BZ}{O}_{SM}
\end{eqnarray}

where BZ, and SM operators are defined as
${O}_{BZ}\in {\cal O}_{BZ}$ with mass dimension $d_{BZ}$,
and ${O}_{SM} \in {\cal O}_{SM} $ with mass dimension $d_{SM}$.
Low energy effects of the scale invariant ${\cal O}_{BZ}$ fields
imply a  dimensional transmutation. Thus, after
the dimensional transmutation Eq.(\ref{1}) is given as

\begin{eqnarray}
\label{2}
 \frac{C_{\cal U} \Lambda_{\cal U}^{d_{BZ}-d}}
{{\cal M}_{\cal U}^k}{O}_{\cal U}{O}_{SM}
\end{eqnarray}

where $d$ is the scaling mass dimension(or anamoulus dimension) 
of the unparticle operator
$O_{\cal U}$ (in Ref.\cite{Georgi:2007ek}, $d=d_{\cal U}$ ),
and the constant $C_{\cal U}$ is a coefficient function.

Interactions between the unparticles and the SM fields 
have been listed by Ref~\cite{Cheung:2007ue}.
Regarding the Georgi's original point of view many work 
on the unparticle physics have been done so far, for example
Ref~\cite{unparticle}.

In this work, we study flavor changing neutral current 
decays $t\to c \gamma$, and $t\to c g$ induced by scalar unparticles.

\section{$t\to c \gamma, g$ decays through unparticle}

The effective interaction between the scalar unparticle and the
SM quarks are given as \cite{Cheung:2007ue}

\begin{equation}
\frac{1}{\Lambda^{d-1}}\bar{f}(\lambda_{S}^{ff'}
+i\gamma_{5}\lambda_{P}^{ff'})f'
\label{eq:3}
\end{equation}

where $f$ and $f'$ denote different flavor of quarks, with
the same electric charge. 

The scalar unparticle propagator is given as

\begin{eqnarray}
 \Delta_{F}(P^{2})=\frac{A_{d}}{2\sin d\pi}(-P^{2}-i\epsilon)^{d-2}
\label{eq:4}
\end{eqnarray}

where

\begin{equation}
A_d=\frac{16\pi^{5/2}}{{(2\pi)}^{2d}}
\frac{\Gamma(d+1/2)}{\Gamma(d-1)\Gamma(2d)}.
\label{eq:5}
\end{equation}

\begin{figure}[h]
\includegraphics
%[height=2cm, width=9cm ]
{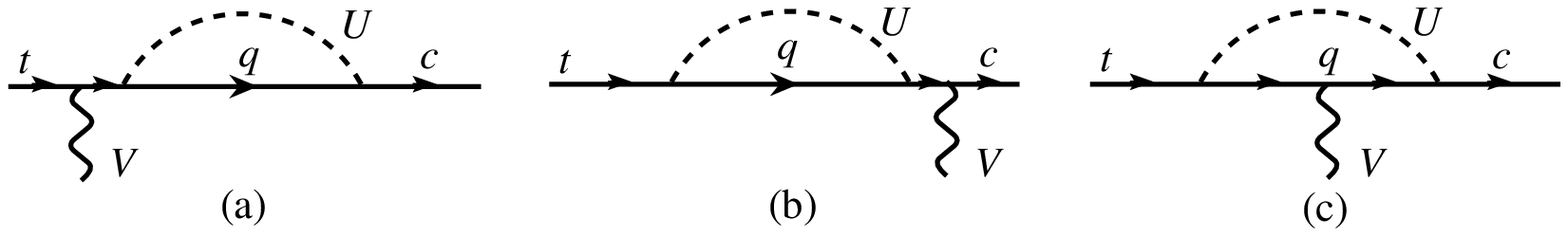}
\caption{Feynman diagrams for FCNC decays of the top quark 
through scalar unparticle.}\label{fig:1}
\end{figure}

The Feynman diagrams for the $t\to c V$ decays through
scalar unparticle is depicted in the Figure~\ref{fig:1}.
The matrix element for the $t\to c V(V=\gamma,g)$ decay in 
general form can be written as follows

\begin{eqnarray}
 M={\epsilon^{\mu }}^{(a) *}
\bar u(p')[i\sigma_{\mu\nu}q^\nu(A_S+A_P \gamma_5)
+\gamma_\mu(C+D\gamma_5)+q_\mu(E+F\gamma_5)]u(p)
\label{eq:6}
\end{eqnarray}
 
where ${\epsilon^{\mu}}^{(a)}$, and $q_\mu=p_\mu-p'_\mu$ are
the polarization, and the momentum vector of the photon(gluon),
respectively, and  $A_S,A_P,C,D,E$ and $F$ are invariant amplitudes.
From the gauge invariance we have $C=D=0$.
Since the photon(gluon) is on shell, i.e. $q^2=0$, and the transversality
condition $q_\mu \epsilon^\mu=0$, leads that the last term in
Eq~(\ref{eq:6}) can safely be omitted. Other words, the $t\to cV$
decay is described by magnetic moment type transition 

\begin{eqnarray}
 M={\epsilon^{\mu }}^{(a) *}
\bar u(p')[i\sigma_{\mu\nu}q^\nu(A_S+A_P \gamma_5)] u(p)
\label{eq:7}
\end{eqnarray}

Obviously the contribution of Fig. 1(a), and Fig. 1(b) are proportional
to ${\epsilon^{\mu }}^{(a) *}\bar u(p')\gamma_\mu u(p)$ or
${\epsilon^{\mu }}^{(a) *}\bar u(p')\gamma_\mu\gamma_5 u(p)$,
and therefore, can be omitted since they do not contribute to
the structure $~\sigma_{\mu\nu}q^\nu$. So, only diagram (c) 
presented in Fig. 1 should be considered. After some calculation 
for the invariant amplitudes $A_{S}$ and $A_{P}$ we get

\begin{eqnarray}
A_{S} & = & \frac{A_d {g}^V}{2\Lambda^{2(d-1)}\sin{d\pi}} 
\sum_{q}  \frac{1}{(4\pi)^{2}}
\int_{0}^{1}dx\int_{0}^{1-x}dy(1-x-y)^{1-d}
%\label{eq:66}
\nonumber\\
 &  & \left[-m_{t}y(1-x-y)(\lambda_{S}^{cq}\lambda_{S}^{tq}
+\lambda_{P}^{cq}\lambda_{P}^{tq})-m_{q}(x+y)(\lambda_{S}^{cq}
\lambda_{S}^{tq}-\lambda_{P}^{cq}\lambda_{P}^{tq})\right]
\nonumber\\
 &  & \left[-(p'x+py)^{2}+p'^{2}x+p^{2}y-m_{q}^{2}(x+y)\right]^{d-2}
\end{eqnarray}

\begin{eqnarray}
A_{P} & = & \frac{A_d {g}^V}{2\Lambda^{2(d-1)}\sin{d\pi}}
\sum_{q}
\frac{1}{(4\pi)^{2}}\int_{0}^{1}dx\int_{0}^{1-x}dy(1-x-y)^{1-d}
%\label{eq:67}
\nonumber\\
 &  &
 \left[-m_{t}y(1-x-y)(-\lambda_{S}^{cq}\lambda_{P}^{tq}+\lambda_{P}^{cq}
\lambda_{S}^{tq})-m_{q}(x+y)(\lambda_{S}^{cq}
\lambda_{P}^{tq}+\lambda_{P}^{cq}\lambda_{S}^{tq})\right]
\nonumber\\
 &  & \left[-(p'x+py)^{2}+p'^{2}x+p^{2}y-m_{q}^{2}(x+y)\right]^{d-2}
\end{eqnarray}

where $q=\{u,c,t\}$, 
when t or c quark running at loop only one of vertices contain  
flavor changing and another vertex is flavor diagonal. But when 
u quark runns at loop both vertices are flavor chaning and 
therefore its contribution to the considered process compared to 
the c and t quark contributions should be very small. For this 
reason we will neglect u quark contributions in all 
next discussions. $\lambda_{S}$($\lambda_{P}$)
and $\Lambda$ are the scalar(pseudo-scalar) couplings and energy
scale of unparticles, respectively. The couplings for the vector bosons
are defined as $g_{}^{\gamma}=Qg_{e}$, 
$g_{}^{g}=g_{s}\lambda^{a}/2$. 

Taking the square and the average of the amplitude gives 

\begin{eqnarray}
 <|M_{c}|^{2}>=
\frac{A_{d}^{2}{g_{}^{V}}^{2}N}{2\Lambda_{}^{4d-4}\sin^{2}d\pi}
\left(|A_{S}|^{2}+|A_{P}|^{2}\right)\left[(p\cdot q)(p'\cdot q)\right]
\end{eqnarray}

where $N$ is color factor given by $\frac{4}{3}$ for the 
$t\to cg$ and $1$ for the $t\to c\gamma$ decay. Therefore, 
the FCNC decay width can be written as 

\begin{eqnarray}
\Gamma^{} = 
\frac{A_{d}^{2}{{g_{}^{V}}^{2}N}(|A_{S}|^{2}+|A_{P}|^{2})}
{32\pi\Lambda_{}^{4d-4}\sin^{2}d\pi}m_{t}^{3}
\end{eqnarray}

The FCNC top quark decay width $\Gamma(t\to Vc)$ is calculated in
terms of the unparticle coupling to the quarks $\lambda$, 
the unparticle scale $\Lambda$ and the scaling dimension $d$. In numerical analysis, without loss of generality, for simplicity, we take 
$\lambda\equiv \lambda_{S}^{(t,c)q}=\lambda_{P}^{(t,c)q}$, and
$m_t=175$ GeV, $m_c=1.2$ GeV, and $\alpha=1/128$, 
$\alpha _s=0.117$. We consider the total width of the top quark decay as $\Gamma_{tot}=1.5$ GeV, which is mainly determined by the decay 
width of $t\to b W^+$, Ref.~\cite{Yao:2006px}.

\begin{figure}[htp!]
\bigskip
\includegraphics
%[height=6cm, width=8cm ]
{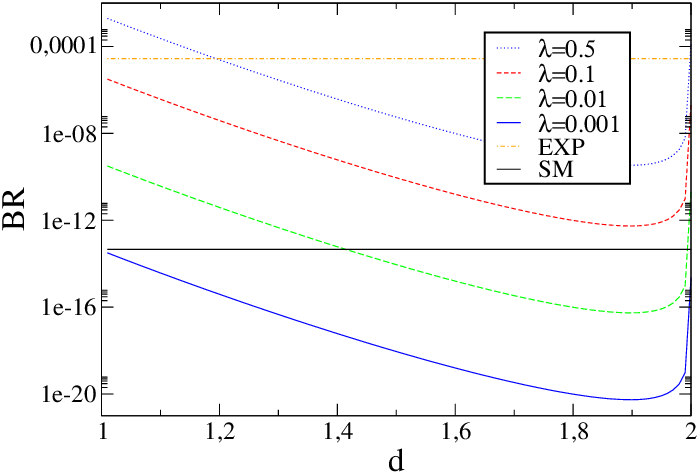} \caption{\label{fig:2}Branching ratio for 
$t\to c\gamma$ decay with $\Lambda=1$TeV.}
\end{figure}

\begin{figure}[hbp!]
\bigskip
\includegraphics
%[height=6cm, width=8cm ]
{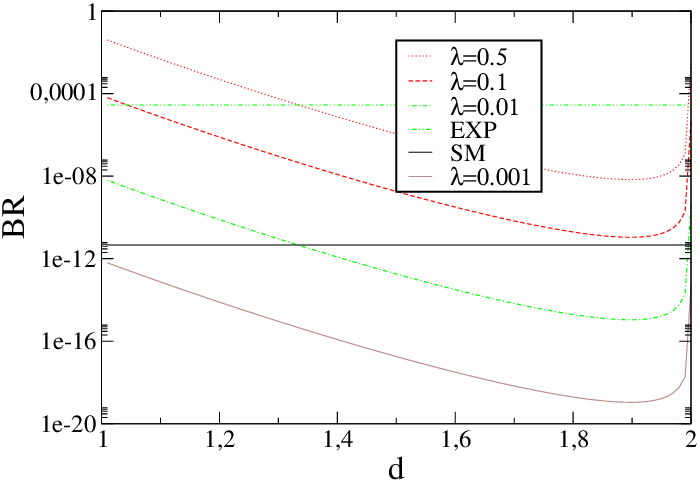} \caption{\label{fig:22}Branching 
ratio for $t\to c g$ decay with  $\Lambda=1$TeV.}
\end{figure}

In Fig. \ref{fig:2}, and Fig. \ref{fig:22}, 
we present the branching ratios 
for $t\to c \gamma$ and $t\to c g$ decays
with respect to the scaling dimension $d$ 
for various values of the coupling $\lambda$ 
at $\Lambda=1$TeV. In these and the following figures
the line (EXP) means 
the result of the simulations performed by the ATLAS Collaboration 
where the upper limits of the considered decays are obtained
about  $10^{-5}$ at $95\%$ C.L., Ref.~\cite{Carvalho:2007yi}.
The SM prediction is represented by the solid horizontal 
line. From those figures we see that the branching ratio of 
for $t\to c \gamma$ and $t\to c g$ decays decreases strongly with
increasing $d$, except $d=2$. It is well known that for 
scalar unparticles at $d=2$ there is infrared singularity.
From the figures it also follows that the branching ratio of
$t\to c \gamma$($t\to c g$) decay becomes smaller than the SM 
prediction when $d\le 1.4$ at $\lambda=10^{-2}$. If the coupling 
constant is larger than $10^{-2}$ then practically at all 
values of $d$ in the considered region
$1<d<2$ branching ratio of $t\to c \gamma$($t\to c g$) decay
in the unparticle theory exceeds the SM one. It should be noted 
that the similar analysis for $b\to s\gamma$, Ref.~\cite{He:2008xv}
(and $\mu\to e \gamma$ decay Ref.~\cite{Ding:2008zza}) 
leads to result that the preferable value of the coupling 
constant is about $\sim 10^{-2}-10^{-3}$ .

\bigskip 

\begin{figure}[htp!]
\bigskip
\includegraphics
%[height=6cm, width=8cm ]
{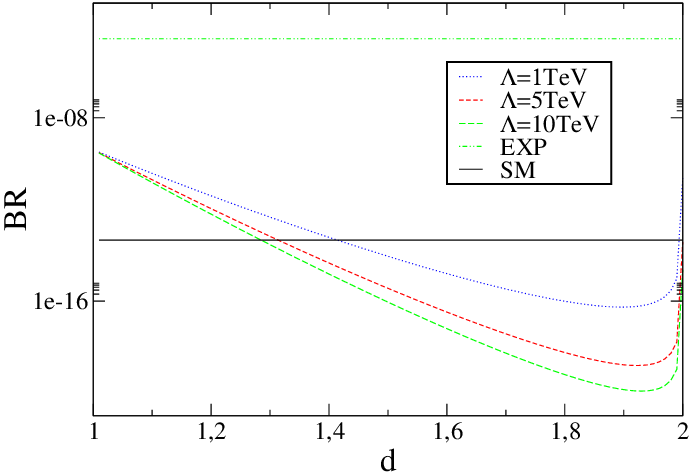} \caption{\label{fig:3}
Branching ratio for $t\to c\gamma$ decay for$\lambda=10^{-2}$.}
\end{figure}

\begin{figure}[hbp!]
\bigskip
\includegraphics
%[height=6cm, width=8cm ]
{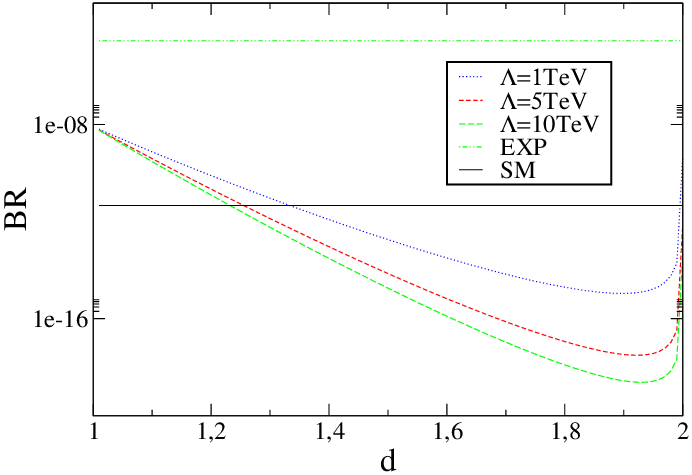} \caption{\label{fig:33} Branching 
ratio for $t\to c g$ decay for $\lambda=10^{-2}$.}
\end{figure}

In the Figures \ref{fig:3}, and \ref{fig:33}
we present the dependence of the branching ratios on 
the parameter $d$ for various values of the energy 
scale $\Lambda$ at $\lambda=10^{-2}$. From these figures
it follows that for $\Lambda=1$TeV - $\Lambda=10$TeV up to $d=1.4$ 
the branching ratio of $t\to c\gamma(t\to cg)$ decay exceeds the SM one.

\bigskip

\bigskip

\begin{figure}[htp!]
\bigskip
\includegraphics
%[height=6cm, width=8cm ]
{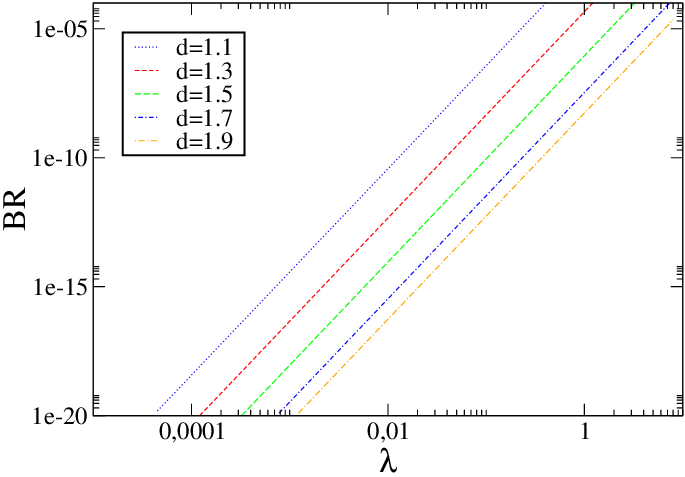} \caption{\label{fig:4}
Branching ratio for $t\to c\gamma$ decay with $\Lambda=1$TeV.}
\end{figure}

\begin{figure}[hbp!]
\bigskip
\bigskip
\includegraphics
%[height=6cm, width=8cm ]
{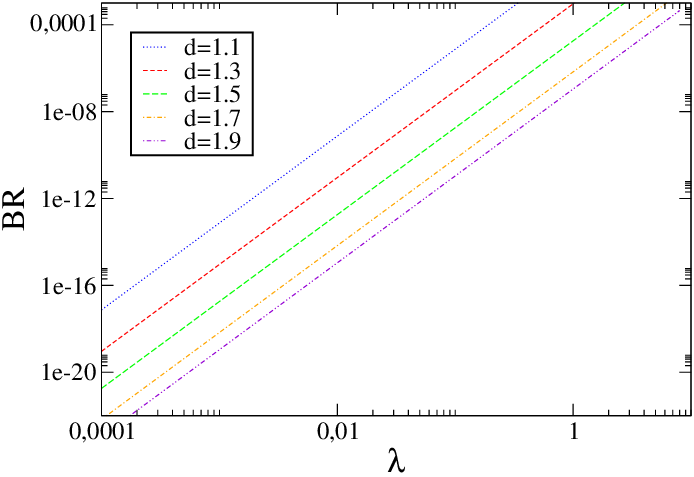} \caption{\label{fig:44}Branching ratio for $t\to c g$ decay with $\Lambda=1$TeV.}
\end{figure}

In the Figure \ref{fig:4}, and \ref{fig:44}
 we present the dependence 
of the branching ratios to the coupling parameter $\lambda$
for given values of the parameter $d$.  From these figures,
one can observe that the branching ratio exceeds the SM prediction
if $\lambda > 10^{-2}$. 

\bigskip

\bigskip

In the Tables \ref{tab1}, and \ref{tab11} we present numerical values
of the branching ratios for  $t\to c g$, and $t\to c\gamma$, respectively.
One can explicitly see that experimental sensitivity is appropriate 
for only $d<1.3$ for $\lambda>1\times 10^{-1}$, however if the 
experimental sensitivity can be increased then the unparticle effects 
can be detected even if the coupling is about $10^{-2}$.

\begin{table}
{\caption{The branching ratio for $t\to c g$ with respect to 
the scaling parameter $d$. Here, we assume $\Lambda=1000$GeV.
\label{tab1}}}
\begin{ruledtabular}
\begin{tabular}{rll}
d & $Br$ for $\lambda=5\times10^{-1}$ & $Br$ for $\lambda=10^{-2}$\\
\hline 
1.1 & $4.7\times 10^{-3}$ & $7.5\times 10^{-10}$ \\
1.2 & $4.9\times 10^{-4}$ & $7.8\times 10^{-11}$ \\
1.3 & $5.7\times 10^{-5}$ & $9.2\times 10^{-12}$ \\
1.4 & $7.5\times 10^{-6}$ & $1.2\times 10^{-12}$ \\
1.5 & $1.1\times 10^{-6}$ & $1.8\times 10^{-13}$ \\
1.6 & $2.0\times 10^{-7}$ & $3.2\times 10^{-14}$ \\
1.7 & $4.2\times 10^{-8}$ & $6.8\times 10^{-15}$ \\
1.8 & $1.2\times 10^{-8}$ & $2.0\times 10^{-15}$ \\
1.9 & $6.8\times 10^{-9}$ & $1.1\times 10^{-15}$ \\
\end{tabular}
\end{ruledtabular}
\end{table}

\bigskip

\begin{table}
{\caption{The branching ratio for $t\to c\gamma$ with respect to 
the scaling parameter $d$. Here, we assume $\Lambda=1000$GeV.
\label{tab11}}}
\begin{ruledtabular}
\begin{tabular}{rll}
d & $Br$ for $\lambda=5\times 10^{-1}$ & $Br$ for $\lambda=10^{-2}$\\
\hline 
1.1 & $2.3\times 10^{-4}$ & $3.7\times 10^{-11}$ \\
1.2 & $2.5\times 10^{-5}$ & $3.9\times 10^{-12}$ \\
1.3 & $2.9\times 10^{-6}$ & $4.6\times 10^{-13}$ \\
1.4 & $3.8\times 10^{-7}$ & $6.1\times 10^{-14}$ \\
1.5 & $5.7\times 10^{-8}$ & $9.0\times 10^{-15}$ \\
1.6 & $9.9\times 10^{-9}$ & $1.6\times 10^{-15}$ \\
1.7 & $2.1\times 10^{-9}$ & $3.4\times 10^{-16}$ \\
1.8 & $6.2\times 10^{-10}$ & $9.7\times 10^{-17}$ \\
1.9 & $3.4\times 10^{-10}$ & $5.5\times 10^{-17}$ \\
\end{tabular}
\end{ruledtabular}
\end{table}

\section{conclusions}

In present work, we study the FCNC rare decays of the top quark 
$t\to c\gamma $ and $t\to c g $ through scalar unparticle.  Regarding the 
latest simulation performed by the ATLAS Collaboration, 
Ref~\cite{Carvalho:2007yi}, the sensitivity to these rare decays of the 
top quark at $\% 95$ C.L. are 
${ Br}(t\to c\gamma)=2.8\times 10^{-5}$, and
${ Br}(t\to c g)=1.6\times 10^{-5}$. 
If there is such a rare decay it will give a window to see the beyond 
SM physics effects. Using the low energy effective field description 
of the unparticle physics we show that FCNC decay of the top quark 
is very good channel to explore for and to put limits on 
the unparticle effects.
We use the limits given by the ATLAS Collaboration to 
constrain the unparticle parameters. According to our results, 
one could expect to see unparticle effects for $\Lambda=1-10$TeV 
if the coupling is about $\lambda > 10^{-2}$ for $d<1.3$. 
This is consistent with the exsisting results in the literature 
(see, for example see Ref.s \cite{unparticle},
\cite{He:2008xv}, \cite{Ding:2008zza}, \cite{Aliev:2008de},  and 
references there in.). 

\bigskip

\begin{table}[htp!]
{\caption{Comparison of the branching ratios
${Br}(t\to c\gamma)$, and ${Br}(t\to c g)$ with the
branching ratios found in the Ref~\cite{Aliev:2008de} for various
values of the scaling papameter $d$. 
\label{tab3}}}
\begin{ruledtabular}
\begin{tabular}{rll}
d & $R_1=\frac{Br(t\to cg)}{Br(t\to c g g)}$ & 
$R_2=\frac{Br(t\to c\gamma)}{Br(t\to c\gamma\gamma)}$\\
\hline 
1.05 & $0.06$ & $0.06$ \\
1.10 & $0.05$ & $0.05$ \\
1.15 & $0.05$ & $0.05$ \\
1.20 & $0.03$ & $0.03$ \\
\end{tabular}
\end{ruledtabular}
\end{table}

We want to remark that
$t\to c\gamma$ or $t\to c g$ are loop level processes both in the SM 
and in the unparticle physics. However, $t\to c \gamma\gamma$ or
$t\to c g g$ can take place at tree level in the unparticle physics   
The unparticle effects in the rare $t\to c g g$ decays has been studied
in the Ref.~\cite{Aliev:2008de}. In Table~\ref{tab3},
we present a comparison our branching ratios ${Br}(t\to c\gamma)$, 
and ${Br}(t\to c g)$ with the branching ratios found 
in the Ref~\cite{Aliev:2008de} for various
values of the scaling parameter $d$. One could understand
this behavior with the observation that the  $t\to c\gamma$ or $t\to c g$
decays are proportional with $\alpha_{em}$ or $\alpha_s$ but the
$t\to c \gamma\gamma$ or $t\to c g g$ decays depend on
the unparticle coupling $\lambda$ which we take $10^{-2}$, is smaller than 
$\alpha_s$ but bigger than $\alpha_{em}$. Therefore, the behaviors of the
branching ratios of $t\to c\gamma(g)$, and $t\to c \gamma\gamma(gg)$ 
in the SM, and the unparticle physics are different.

\begin{acknowledgments}
OC acknowledges the support from CERN PH Department.
The work of OC was supported in part by
the State Planning Organization (DPT) under grant no DPT-2006K-120470.
\end{acknowledgments}


\begin{thebibliography}{1}

%\cite{Beneke:2000hk}
\bibitem{Beneke:2000hk}
  M.~Beneke {\it et al.},
  %``Top quark physics,''
  arXiv:hep-ph/0003033.
  %%CITATION = HEP-PH/0003033;%%

%\cite{Han:2008xb}
\bibitem{Han:2008xb}
  T.~Han,
  %``The 'Top Priority' at the LHC,''
  arXiv:0804.3178 [hep-ph].
  %%CITATION = ARXIV:0804.3178;%%

%\cite{Bernreuther:2008ju}
\bibitem{Bernreuther:2008ju}
  W.~Bernreuther,
  %``Top quark physics at the LHC,''
  J.\ Phys.\ G {\bf 35}, 083001 (2008)
  [arXiv:0805.1333 [hep-ph]].
  %%CITATION = JPHGB,G35,083001;%%


\bibitem{Yao:2006px} W.~M.~Yao \textit{et al.} {[}Particle Data
Group], J.\ Phys.\ G \textbf{33}, 1 (2006). 


%\cite{Chakraborty:2003iw}
\bibitem{Chakraborty:2003iw}
  D.~Chakraborty, J.~Konigsberg and D.~L.~Rainwater,
  %``Review of top quark physics,''
  Ann.\ Rev.\ Nucl.\ Part.\ Sci.\  {\bf 53}, 301 (2003)
  [arXiv:hep-ph/0303092].
  %%CITATION = ARNUA,53,301;%%

%\cite{Eilam:1990zc}
\bibitem{Eilam:1990zc} 
  G.~Eilam, J.~L.~Hewett and A.~Soni,
  %``Rare decays of the top quark in the standard and two Higgs doublet
  %models,''
  Phys.\ Rev.\  D {\bf 44}, 1473 (1991)
  [Erratum-ibid.\  D {\bf 59}, 039901 (1999)].
  %%CITATION = PHRVA,D44,1473;%%


%\cite{AguilarSaavedra:2004wm}
\bibitem{AguilarSaavedra:2004wm}
  J.~A.~Aguilar-Saavedra,
  %``Top flavour-changing neutral interactions: Theoretical expectations and
  %experimental detection,''
  Acta Phys.\ Polon.\  B {\bf 35}, 2695 (2004)
  [arXiv:hep-ph/0409342].
  %%CITATION = APPOA,B35,2695;%%

%\cite{Frank:2005vd}
\bibitem{Frank:2005vd}
  M.~Frank and I.~Turan,
  %``t --> c g, c gamma, c Z in the left-right supersymmetric model,''
  Phys.\ Rev.\  D {\bf 72}, 035008 (2005)
  [arXiv:hep-ph/0506197].
  %%CITATION = PHRVA,D72,035008;%%

%\cite{Mele:1998ag}
\bibitem{Mele:1998ag}
  B.~Mele, S.~Petrarca and A.~Soddu,
  %``A new evaluation of the t --> c H decay width in the standard model,''
  Phys.\ Lett.\  B {\bf 435}, 401 (1998)
  [arXiv:hep-ph/9805498].
  %%CITATION = PHLTA,B435,401;%%

%\cite{Carvalho:2007yi}
\bibitem{Carvalho:2007yi}
  J.~Carvalho {\it et al.}  [ATLAS Collaboration],
  %``Study of ATLAS sensitivity to FCNC top decays,''
  Eur.\ Phys.\ J.\  C {\bf 52}, 999 (2007)
  [arXiv:0712.1127 [hep-ex]];
ATLAS Collaboration, Expected Performance of the ATLAS Experiment,
Detector, Trigger and Physics, CERN-OPEN-2008-020, Geneva, 2008, to
appear. 
  %%CITATION = EPHJA,C52,999;%%

%\cite{Han:1995pk}
\bibitem{Han:1995pk}
  T.~Han, R.~D.~Peccei and X.~Zhang,
  %``Top Quark Decay Via Flavor Changing Neutral Currents At Hadron Colliders,''
  Nucl.\ Phys.\  B {\bf 454} (1995) 527
  [arXiv:hep-ph/9506461].
  %%CITATION = NUPHA,B454,527;%%

%\cite{Li:1993mg}
\bibitem{Li:1993mg}
  C.~S.~Li, R.~J.~Oakes and J.~M.~Yang,
  %``Rare decay of the top quark in the minimal supersymmetric model,''
  Phys.\ Rev.\  D {\bf 49}, 293 (1994)
  [Erratum-ibid.\  D {\bf 56}, 3156 (1997)].
  %%CITATION = PHRVA,D49,293;%%

%\cite{deDivitiis:1997sh}
\bibitem{deDivitiis:1997sh}
  G.~M.~de Divitiis, R.~Petronzio and L.~Silvestrini,
  %``Flavour changing top decays in supersymmetric extensions of the  standard
  %model,''
  Nucl.\ Phys.\  B {\bf 504}, 45 (1997)
  [arXiv:hep-ph/9704244].
  %%CITATION = NUPHA,B504,45;%%

%\cite{Lopez:1997xv}
\bibitem{Lopez:1997xv}
  J.~L.~Lopez, D.~V.~Nanopoulos and R.~Rangarajan,
  %``New supersymmetric contributions to t --> c V,''
  Phys.\ Rev.\  D {\bf 56}, 3100 (1997)
  [arXiv:hep-ph/9702350].
  %%CITATION = PHRVA,D56,3100;%%

%\cite{HongSheng:2007ve}
\bibitem{HongSheng:2007ve}
  H.~Hong-Sheng,
  %``Flavor-changing top quark rare decays in the littlest Higgs model with
  %T-parity,''
  Phys.\ Rev.\  D {\bf 75}, 094010 (2007)
  [arXiv:hep-ph/0703067].
  %%CITATION = PHRVA,D75,094010;%%

%%%%%%%%%%%%%%%%%%%%%%%%%%%%%%%%%%%

\bibitem{Georgi:2007ek} H.~Georgi, Phys.\ Rev.\ Lett.\ \textbf{98},
221601 (2007) {[}arXiv:hep-ph/0703260]. 

\bibitem{Georgi:2007si} H.~Georgi, Phys.\ Lett.\ B \textbf{650},
275 (2007) {[}arXiv:0704.2457 {[}hep-ph]]. 


\bibitem{Banks:1981nn} T.~Banks and A.~Zaks, Nucl.\ Phys.\ B
\textbf{196}, 189 (1982). 

%%%%%%
%\cite{Cheung:2007ue}
\bibitem{Cheung:2007ue}
  K.~Cheung, W.~Y.~Keung and T.~C.~Yuan,
  %``Novel signals in unparticle physics,''
  Phys.\ Rev.\ Lett.\  {\bf 99}, 051803 (2007)
  arXiv:0704.2588 [hep-ph];
  %%CITATION = ARXIV:0704.2588;%%
  %``Collider Phenomenology of Unparticle Physics,''
  Phys.\ Rev.\  D {\bf 76}, 055003 (2007)
  [arXiv:0706.3155 [hep-ph]].
  %%CITATION = PHRVA,D76,055003;%%
S.~L.~Chen, X.~G.~He and H.~C.~Tsai,
  %``Constraints on Unparticle Interactions from Invisible Decays of Z,
  %Quarkonia and Neutrinos,''
  arXiv:0707.0187 [hep-ph].
  %%CITATION = ARXIV:0707.0187;%%
\bibitem{unparticle}
%\cite{Aliev:2007qw}
%\bibitem{Aliev:2007qw}
  T.~M.~Aliev, A.~S.~Cornell and N.~Gaur,
  %``Lepton Flavour Violation in Unparticle Physics,''
  Phys.\ Lett.\  B {\bf 657}, 77 (2007)
  [arXiv:0705.1326 [hep-ph]].
  %%CITATION = PHLTA,B657,77;%%
%\cite{Bander:2007nd}
%\bibitem{Bander:2007nd}
  M.~Bander, J.~L.~Feng, A.~Rajaraman and Y.~Shirman,
  %``Unparticles: Scales and High Energy Probes,''
  Phys.\ Rev.\  D {\bf 76}, 115002 (2007)
  [arXiv:0706.2677 [hep-ph]];
  %%CITATION = PHRVA,D76,115002;%%
%\cite{Davoudiasl:2007jr}
%\bibitem{Davoudiasl:2007jr}
  H.~Davoudiasl,
  %``Constraining Unparticle Physics with Cosmology and Astrophysics,''
  Phys.\ Rev.\ Lett.\  {\bf 99}, 141301 (2007)
  [arXiv:0705.3636 [hep-ph]];
  %%CITATION = PRLTA,99,141301;%%
%\cite{Ding:2007bm}
%\bibitem{Ding:2007bm}
  G.~J.~Ding and M.~L.~Yan,
  %``Unparticle Physics in DIS,''
  Phys.\ Rev.\  D {\bf 76}, 075005 (2007)
  [arXiv:0705.0794 [hep-ph]];
  %%CITATION = PHRVA,D76,075005;%%
%\cite{Aliev:2007gr}
%\bibitem{Aliev:2007gr}
  T.~M.~Aliev, A.~S.~Cornell and N.~Gaur,
  %``B \to K(K^*) missing energy in Unparticle physics,''
  JHEP {\bf 0707}, 072 (2007)
  [arXiv:0705.4542 [hep-ph]].
  %%CITATION = JHEPA,0707,072;%%
%\cite{Balantekin:2007eg}
%\bibitem{Balantekin:2007eg}
  A.~B.~Balantekin and K.~O.~Ozansoy,
  %``Constraints on Unparticles from Low Energy Neutrino-Electron Scattering,''
  arXiv:0710.0028 [hep-ph].
  %%CITATION = ARXIV:0710.0028;%%
%\cite{Aliev:2007rm}
%\bibitem{Aliev:2007rm}
  T.~M.~Aliev and M.~Savci,
  %``Unparticle physics effects in (Lambda_b -> Lambda + missing energy)
  %processes,''
  Phys.\ Lett.\  B {\bf 662}, 165 (2008)
  [arXiv:0710.1505 [hep-ph]].
  %%CITATION = PHLTA,B662,165;%%
%\cite{Rizzo:2007xr}
%\bibitem{Rizzo:2007xr}
  T.~G.~Rizzo,
  %``Contact Interactions and Resonance-Like Physics at Present and Future
  %Colliders from Unparticles,''
  JHEP {\bf 0710}, 044 (2007)
  [arXiv:0706.3025 [hep-ph]];
  %%CITATION = JHEPA,0710,044;%%
%\cite{Cakir:2007xb}
%\bibitem{Cakir:2007xb}
  O.~Cakir and K.~O.~Ozansoy,
  %``Unparticle Searches Through Gamma Gamma Scattering,''
  Eur.\ Phys.\ J.\  C {\bf 56}, 279 (2008)
  [arXiv:0712.3814 [hep-ph]].
  %%CITATION = EPHJA,C56,279;%%
%\cite{Cakir:2007dz}
%\bibitem{Cakir:2007dz}
  O.~Cakir and K.~O.~Ozansoy,
  %``Unparticle Searches Through Compton Scattering,''
 Eur.\ Phys.\ Lett. {\bf 83}, 51001 (2008)  
[arXiv:0710.5773 [hep-ph]].
  %%CITATION = ARXIV:0710.5773;%%
%\cite{Fox:2007sy}
%\bibitem{Fox:2007sy}
  P.~J.~Fox, A.~Rajaraman and Y.~Shirman,
  %``Bounds on Unparticles from the Higgs Sector,''
  Phys.\ Rev.\  D {\bf 76}, 075004 (2007)
  [arXiv:0705.3092 [hep-ph]];
  %%CITATION = PHRVA,D76,075004;%%
%\cite{Liao:2007bx}
%\bibitem{Liao:2007bx}
  Y.~Liao,
  %``Bounds on Unparticles Couplings to Electrons: from Electron g-2 to
  %Positronium Decays,''
  Phys.\ Rev.\  D {\bf 76}, 056006 (2007)
  [arXiv:0705.0837 [hep-ph]];
  %%CITATION = PHRVA,D76,056006;%%
%\cite{Liao:2007ic}
%\bibitem{Liao:2007ic}
  Y.~Liao and J.~Y.~Liu,
  %``Long-range Electron Spin-spin Interactions from Unparticle Exchange,''
  Phys.\ Rev.\ Lett.\  {\bf 99}, 191804 (2007)
  [arXiv:0706.1284 [hep-ph]];
  %%CITATION = PRLTA,99,191804;%%
%\cite{Iltan:2007ve}
%\bibitem{Iltan:2007ve}
  E.~O.~Iltan,
  %``The scalar unparticle effect on the charged lepton electric dipole
  %moment,''
  arXiv:0710.2677 [hep-ph];
  %%CITATION = ARXIV:0710.2677;%%
%\cite{Lenz:2007nj}
%\bibitem{Lenz:2007nj}
  A.~Lenz,
  %``Unparticle physics effects in B_s mixing,''
  Phys.\ Rev.\  D {\bf 76}, 065006 (2007)
  [arXiv:0707.1535 [hep-ph]];
  %%CITATION = PHRVA,D76,065006;%%
%\cite{Alan:2007ss}
%\bibitem{Alan:2007ss}
  A.~T.~Alan and N.~K.~Pak,
  %``Unparticle physics in top pair signals at the LHC and ILC,''
  arXiv:0708.3802 [hep-ph];
  %%CITATION = ARXIV:0708.3802;%%
%%%%%%%%%%%%%%%%%%%%%%%%%%%%%%%%%%%%%%%%%%%%%%%%

%\cite{He:2008xv}
\bibitem{He:2008xv}
  X.~G.~He and L.~Tsai,
  %``Constraints on Unparticle Interaction from $b\to s \gamma$,''
  JHEP {\bf 0806}, 074 (2008)
  [arXiv:0805.3020 [hep-ph]].
  %%CITATION = JHEPA,0806,074;%%

%\cite{Ding:2008zza}
\bibitem{Ding:2008zza}
  G.~J.~Ding and M.~L.~Yan,
  %``Lepton flavor violating mu- $\to$ e gamma and mu-e conversion in unparticle
  %physics,''
  Phys.\ Rev.\  D {\bf 77}, 014005 (2008).
  %%CITATION = PHRVA,D77,014005;%%

%\cite{Aliev:2008de}
\bibitem{Aliev:2008de}
  T.~M.~Aliev, A.~Bekmezci and M.~Savci,
  %``Unparticle effects in rare (t -> c g g) decay,''
  Phys.\ Rev.\ D {\bf 78}, 057701 (2008) 
  [arXiv:0805.1150 [hep-ph]].
  %%CITATION = ARXIV:0805.1150;%%

\end{thebibliography}
\end{document}